# An adaptive multi-level simulation algorithm for stochastic biological systems

C. Lester,[1, a)] C.A. Yates,[2] M.B. Giles,[1] and R.E. Baker[1]

[1)]*Mathematical Institute, Woodstock Road, Oxford, OX2 6GG*

[2)]*Department of Mathematical Sciences, Bath, BA2 7AY*

(Dated: 9 December 2014)

Discrete-state, continuous-time Markov models are widely used in the modeling of biochemical reaction networks. Their complexity often precludes analytic solution, and we rely on stochastic simulation algorithms to estimate system statistics. The Gillespie algorithm is exact, but computationally costly as it simulates every single reaction. As such, approximate stochastic simulation algorithms such as the tau-leap algorithm are often used. Potentially computationally more efficient, the system statistics generated suffer from significant bias unless tau is relatively small, in which case the computational time can be comparable to that of the Gillespie algorithm.

The multi-level method (Anderson and Higham, Multiscale Model. Simul. 10:146–179, 2012) tackles this problem. A base estimator is computed using many (cheap) sample paths at low accuracy. The bias inherent in this estimator is then reduced using a number of corrections. Each correction term is estimated using a collection of paired sample paths where one path of each pair is generated at a higher accuracy compared to the other (and so more expensive). By sharing random variables between these paired paths the variance of each correction estimator can be reduced. This renders the multi-level method very efficient as only a relatively small number of paired paths are required to calculate each correction term.

In the original multi-level method, each sample path is simulated using the tau-leap algorithm with a fixed value of $\tau$. This approach can result in poor performance when the reaction activity of a system changes substantially over the timescale of interest. By introducing a novel, adaptive time-stepping approach where $\tau$ is chosen according to the stochastic behaviour of each sample path we extend the applicability of the multi-level method to such cases. We demonstrate the efficiency of our method using a number of examples.

---

[a)]Electronic mail: lesterc@maths.ox.ac.uk

I. INTRODUCTION

Experimental researchers such as Elowitz et al.[1], Fedoroff and Fontana[2], Arkin et al.[3] and Barrio et al.[4] have demonstrated the stochastic nature of a range of biological phenomena. In particular, stochastic effects often affect systems characterized by low molecular populations[5], but systems with large molecular populations can also be affected under certain circumstances[6]. In this work we will focus on spatially homogeneous population-level models, which record the numbers of each molecule type within the system over a time interval of interest. The temporal evolution of the probability of finding each combination of possible 'molecular abundancies' can be described by the chemical master equation (CME), which comprises a system of ordinary differential equations (ODEs)[5]. Under highly restrictive conditions it is possible to derive a closed-form, analytic solution of the CME[7]. However, in a more typical setting we are restricted to understanding the behavior of a particular system computationally by using a stochastic simulation algorithm (SSA) to generate a large number of sample paths; we then use these paths as a means to calculate ensemble statistics that describe the quantitative behavior of the system. Such SSAs can either be exact or approximate. Exact SSAs generate sample paths consistent with the dynamics of the CME hence give rise to unbiased estimators[8], whereas the sample paths generated using approximate SSAs do not fully comply with the CME and give rise to biased estimators[9].

The focus of this paper is on extending the discrete-state multi-level technique first introduced by Anderson and Higham[10]. Their approach broadly emulates that of Giles[11] in the field of stochastic differential equations. The multi-level method uses a clever combination of sample paths generated using approximate SSAs of different accuracy to estimate a system statistic of interest in an efficient manner[12]. The idea of the multi-level method is to compute many (cheap) sample paths with low accuracy and correct the statistics generated from them using fewer (expensive) sample paths with high accuracy. Each approximate SSA involves using a fixed time discretization: paths that use a fine time discretization are more expensive but display less bias whereas paths with a coarse time discretization are cheap but heavily biased. If properly implemented, use of the multi-level method can often lead to substantial reductions in simulation time. However, the original formulation of the method places restrictions on the time-discretizations implemented in the multi-level method. Specifically

these are:

- that the time discretization be uniform;

- that the time discretizations for sample paths of different accuracy are *nested*. In other words, the time step is reduced by some integer factor $K \in \{2, 3, \ldots\}$ between consecutive accuracy levels.

In this work, we will show that such restrictions mean the multi-level method can be inefficient at simulating systems in which the reaction activity changes significantly over the time interval of interest. Such systems include those that display stiff behavior, where there are markedly different timescales displayed over the course of the simulation. We address this issue with a new *adaptive* approach to the multi-level method: the time discretization for the approximate SSA is chosen on the fly, taking into account the reaction activity within each individual sample path.

The remainder of this paper is structured as follows. In Section II we briefly recapitulate the technical construction of stochastic simulations of discrete-state systems; this is followed by a brief introduction to the multi-level method. Readers seeking a more thorough exposition may be interested in a number of survey papers[5,12,13]. In Section III we highlight two cases where the fixed time-step multi-level method is unable function efficiently and in Section IV we present a novel *adaptive* multi-level method as a solution to this problem. The benefits of our new method are fully explored with reference to the motivating examples in Section V. We conclude by making sense of these results in Section VI, and provide some possibilities for directions of future research based on our findings.

## II. STOCHASTIC SIMULATION

We consider a reaction network comprising $N$ species, $S_1, \ldots, S_N$, that may each be involved in $M$ possible interactions, $R_1, \ldots, R_M$, referred to as reaction channels. For the purpose of this discussion, we will ignore spatial effects. The population size of $S_i$ is known as its copy number and is denoted by $X_i(t)$ at time $t$, $t \geq 0$. The state vector of all species numbers is then defined as

$$\boldsymbol{X}(t) := \begin{bmatrix} X_1(t) \\ \vdots \\ X_N(t) \end{bmatrix}. \qquad (1)$$

With each reaction channel, $R_j$, we associate two quantities. The first is the stoichiometric or state-change vector,

$$\boldsymbol{\nu}_j := \begin{bmatrix} \nu_{1j} \\ \vdots \\ \nu_{Nj} \end{bmatrix}, \qquad (2)$$

where $\nu_{ij}$ is the change in the copy number of $S_i$ caused by reaction $R_j$ taking place. Thus if the system is in state $\boldsymbol{X}$ and reaction $R_j$ happens, the system jumps to state $\boldsymbol{X} + \boldsymbol{\nu}_j$. The second quantity is the propensity function, $a_j$. This represents the rate at which a reaction takes place. Formally, for small $\mathrm{d}t$, and supposing that $\boldsymbol{X}(t) = \boldsymbol{x}$, we define $a_j(\boldsymbol{x})$ as follows:

- the probability that reaction $R_j$ happens exactly once during the infinitesimal interval $[t, t+\mathrm{d}t)$ is $a_j(\boldsymbol{x})\mathrm{d}t + o(\mathrm{d}t)$;

- the probability of more than one reaction $R_j$ during this interval is $o(\mathrm{d}t)$.

Our approach to understanding the dynamics of the system comes from considering how the probability that the system is in a particular state changes through time. Defining

$$\mathbb{P}(\boldsymbol{x}, t \mid \boldsymbol{x_0}, t_0) \equiv \mathbb{P}\left[\boldsymbol{X}(t) = \boldsymbol{x}, \text{ given } \boldsymbol{X}(t_0) = \boldsymbol{x_0}\right],$$

then the CME gives[5]

$$\frac{\mathrm{d}\mathbb{P}(\boldsymbol{x}, t \mid \boldsymbol{x_0}, t_0)}{\mathrm{d}t} = \sum_{j=1}^{M} [\mathbb{P}(\boldsymbol{x} - \boldsymbol{\nu}_j, t \mid \boldsymbol{x_0}, t_0) \cdot a_j(\boldsymbol{x} - \boldsymbol{\nu}_j) - \mathbb{P}(\boldsymbol{x}, t \mid \boldsymbol{x_0}, t_0) \cdot a_j(\boldsymbol{x})]. \qquad (3)$$

The Kurtz representation[14] of a reaction network provides an equivalent, alternative formulation: each reaction channel is described with an inhomogeneous Poisson process, and the dynamics of the CME are preserved.

The simplest, and perhaps most widely used method for generating sample paths in accordance with the CME is Gillespie's Direct Method (DM)[8]. Subsequent work has substantially refined this approach[15–18], and it is against this gold standard that we shall compare the multi-level method.

## A. Tau leaping

Constraints on computing resources often limit the feasibility of the Gillespie DM as it simulates each reaction individually. The large costs in doing so come from two main sources: first is the computational overheads in generating the large quantity of random numbers required by the algorithm; and second is the search time involved in determining which reaction occurs at each step.

The tau-leaping method, first proposed by Gillespie[9], generates approximate sample paths by taking steps, of length $\tau$, throughout the time-interval of interest and firing several reactions during each time step. In this way it 'leaps' over several reactions at a time. If the system is in state $\boldsymbol{X}$ and a time step of $\tau$ is to be performed, let $K_j(\tau, \boldsymbol{X})$ represent the number of times that reaction channel $R_j$ fires within that time step. The key, time-saving assumption of the tau-leaping method is that all reaction rates are assumed to remain constant over each time step of length $\tau$. This means that $K_j(\tau, \boldsymbol{X})$ is Poisson distributed[9], $K_j \sim \text{Poisson}(a_j(\boldsymbol{X}(t)) \cdot \tau)$. The tau-leaping algorithm proceeds at each time step by generating Poisson random variates with the correct parameter for each reaction channel, and then updating each molecular species and propensity function simultaneously:

$$\boldsymbol{X}(t+\tau) = \boldsymbol{X}(t) + \sum_{j=1}^{M} K_j(\tau, \boldsymbol{X}(t))\boldsymbol{\nu}_j. \quad (4)$$

Appropriate choices of $\tau$ must be used throughout the construction of a sample path. Smaller values of $\tau$ will lead to point estimates with a lower bias[9], but will require more steps to simulate a path and therefore a higher run-time.

The core algorithm for the tau-leaping algorithm which terminates at time $T$ proceeds as follows[9]:

1. set $\boldsymbol{X} := \boldsymbol{X}(t_0)$ and $t := t_0$;

2. calculate the leap size, $\tau$. If $t + \tau > T$ then set $\tau := T - t$;

3. calculate the propensity function, $a_j$, for each reaction channel, $R_j, j = 1, \ldots, M$, based on $\boldsymbol{X}(t)$, the population vector at time $t$;

4. generate Poisson random variates, $p_j$, as sample values of $K_j(\tau, \boldsymbol{X}(t)), j = 1, \ldots, M$;

5. set $\boldsymbol{X} := \boldsymbol{X} + \sum_{j=1}^{M} p_j \boldsymbol{\nu}_j$ and $t := t + \tau$;

6. if $t < T$, return to step two.

A range of techniques have been developed to balance the competing priorities of speed and accuracy in choosing $\tau$ in step two[9,19–21].

## B. The multi-level method

The original multi-level method divides the work done in calculating a system statistic of interest into parts, known as levels, in an effort to increase computational efficiency[12]. On each of the levels point estimates are calculated using the tau-leaping algorithm with different, *fixed* values of $\tau$. They are then summed to give the point estimate of interest. Suppose we wish to estimate the expected value of $X_i$, the population of the species $S_i$, at time $T$. On the base level (level 0), a tau-leaping method with a large and fixed value of $\tau$ (which we denote $\tau_0$) is used to generate a large number ($n_0$) of sample paths of the system. The resulting point estimate for $X_i$ is

$$Q_0 := \mathbb{E}\left[Z_{\tau_0}\right] \approx \frac{1}{n_0} \sum_{r=1}^{n_0} Z_{\tau_0}^{(r)}, \tag{5}$$

where $Z_\tau^{(r)}$ is the copy number of $S_i$ (the species of interest) at terminal time $T$ in path $r$ generated using the tau-leaping method with time step $\tau$, and $n_\ell$ is the number of paths generated on level $\ell$. As $\tau$ is large, this estimate is calculated cheaply ($\mathcal{O}(1/\tau)$ units of time are required to generate each sample path), with the downside being that it has considerable bias.

The goal with the next level (level 1) is to introduce a correction term that begins to reduce this bias. In essence, in order to compute this correction term, two sets of $n_1$ sample paths are calculated. One set is generated using the tau-leaping method with the same value of $\tau$ as on the base level ($\tau_0$). The other set is generated using a smaller value of $\tau$ (which we denote $\tau_1$). For the method of Anderson and Higham[10] to work, we require that $\tau_1 = \tau_0/K$, where $K \in \{2, 3, \ldots\}$. The correction term is the difference between the point estimates calculated from each set of sample paths:

$$Q_1 := \mathbb{E}\left[Z_{\tau_1} - Z_{\tau_0}\right] \approx \frac{1}{n_1} \sum_{r=1}^{n_1} \left[Z_{\tau_1}^{(r)} - Z_{\tau_0}^{(r)}\right]. \tag{6}$$

Adding this correction term to the estimator calculated on the base level reduces the bias of the resulting estimator. This can be seen by noting that $Q_0 + Q_1 = \mathbb{E}[Z_{\tau_0}] + \mathbb{E}[Z_{\tau_1} - Z_{\tau_0}] = \mathbb{E}[Z_{\tau_1}]$, so that the sum of the two estimators has a bias equivalent to that of the tau-leaping method with $\tau = \tau_1$. The key to the efficiency of the multi-level method is to generate the two sets of sample paths,

$$\left\{ Z_{\tau_1}^{(r)}, \, Z_{\tau_0}^{(r)} : r = 1, \ldots, n_1 \right\}, \tag{7}$$

in a clever way, so that the variance in their difference is minimised. On the next level (level 2), this process is repeated to give a second correction term. Two sets of $n_2$ sample paths are generated, one set has $\tau = \tau_1$, and the second has $\tau = \tau_2 = \tau_1/K = \tau_0/K^2$. Again, the correction term is the estimator of their difference,

$$Q_2 := \mathbb{E}[Z_{\tau_2} - Z_{\tau_1}] \approx \frac{1}{n_2} \sum_{r=1}^{n_2} \left[ Z_{\tau_2}^{(r)} - Z_{\tau_1}^{(r)} \right], \tag{8}$$

and it is added to the combined estimator from level 0 and level 1 to give $\mathbb{Q} = Q_0 + Q_1 + Q_2 = \mathbb{E}[Z_{\tau_2}]$. Carrying on in this way, the multi-level method forms a telescoping sum,

$$\mathbb{Q} = \mathbb{E}[Z_{\tau_L}] = \mathbb{E}[Z_{\tau_0}] + \sum_{\ell=1}^{L} \mathbb{E}[Z_{\tau_\ell} - Z_{\tau_{\ell-1}}] = \sum_{\ell=0}^{L} Q_\ell. \tag{9}$$

With the addition of each subsequent level the bias of the estimator is reduced further, until a desired level of accuracy is reached.

Finally, and optionally, by generating two sets of $n_{L+1}$ sample paths, one set using an exact SSA and the other using tau-leaping with $\tau = \tau_L$, we can efficiently compute a final correction term,

$$Q_{L+1}^* = \mathbb{E}[X_i - Z_{\tau_L}] \approx \frac{1}{n_{L+1}} \sum_{r=1}^{n_{L+1}} \left[ X_i^{(r)} - Z_{\tau_L}^{(r)} \right], \tag{10}$$

where $X_i^{(r)}$ denotes the copy number of $S_i$ at terminal time $T$ in path $r$ generated using the exact SSA. This final correction term can be added to the telescoping sum in order to make the estimator unbiased:

$$\mathbb{Q} = \mathbb{E}[X_i] = \mathbb{E}[Z_{\tau_0}] + \sum_{\ell=1}^{L} \mathbb{E}[Z_{\tau_\ell} - Z_{\tau_{\ell-1}}] + \mathbb{E}[X_i - Z_{\tau_L}] = \sum_{\ell=0}^{L} Q_\ell + Q_{L+1}^*. \tag{11}$$

Importantly, the total time taken to generate the sets of sample paths for the base level, $Q_0$, and each of the correction terms, $Q_\ell$ for $\ell = 1, \ldots, L$, and $Q_{L+1}^*$, can be less than that taken to estimate $\mathbb{E}[X_i]$ using an exact SSA. In order for these time savings to be realized, however, the algorithm needs to be carefully calibrated. This involves choosing the number of levels and the time steps involved (governed by a choice of a scaling factor $K$ and $L$); further details (including a method for choosing $\tau_0$) are given in Lester et al.[22].

Having decided on these parameters, we turn to determining the number of sample paths which need to be simulated for each level $\ell$. If each sample path on level $\ell$ takes $c_\ell$ units of time to produce, then we choose the number of paths, $n_\ell$, such that the total run time (over all levels), $\sum_{\ell \geq 0} c_\ell \cdot n_\ell$, is minimized, subject to the constraint that the overall estimator variance is bounded by some $\varepsilon$. If level $\ell$ has a sample variance of $\sigma_\ell^2$, this makes its estimator variance $\sigma_\ell^2/n_\ell$, and the overall estimator variance $\sum_{\ell \geq 0} \sigma_\ell^2/n_\ell$. Hence this constraint can be expressed as $\sum_{\ell \geq 0} \sigma_\ell^2/n_\ell < \varepsilon$. Generally, $\sigma_\ell^2$ cannot be calculated *a priori* and estimates based on a small number of preliminary simulations, which are produced as part of a model calibration procedure, can be used instead[10].

In many circumstances the multi-level method can provide significant computational savings[10,12,22]. However, it is possible that for systems in which the reaction activity changes significantly over the time scale of interest the efficiency of the fixed time step multi-level approach will be limited. In the next section we consider two reaction systems in which the fixed time-step multi-level approach provides only a limited degree of acceleration over Gillespie's DM.

## III. TWO MOTIVATING EXAMPLES

This section introduces two motivating examples which highlight potential limitations of the fixed time-step multi-level method.

### A. Case Study I: A dimerization model

This following system has been employed widely as a test of stochastic simulation algorithms[9,23] as it exhibits behaviors on multiple timescales. The reaction network is given by:

$$R_1 : S_1 \xrightarrow{1} \emptyset; \qquad R_2 : S_2 \xrightarrow{1/25} S_3; \\ R_3 : S_1 + S_1 \xrightarrow{1/500} S_2; \qquad R_4 : S_2 \xrightarrow{1/2} S_1 + S_1. \tag{12}$$

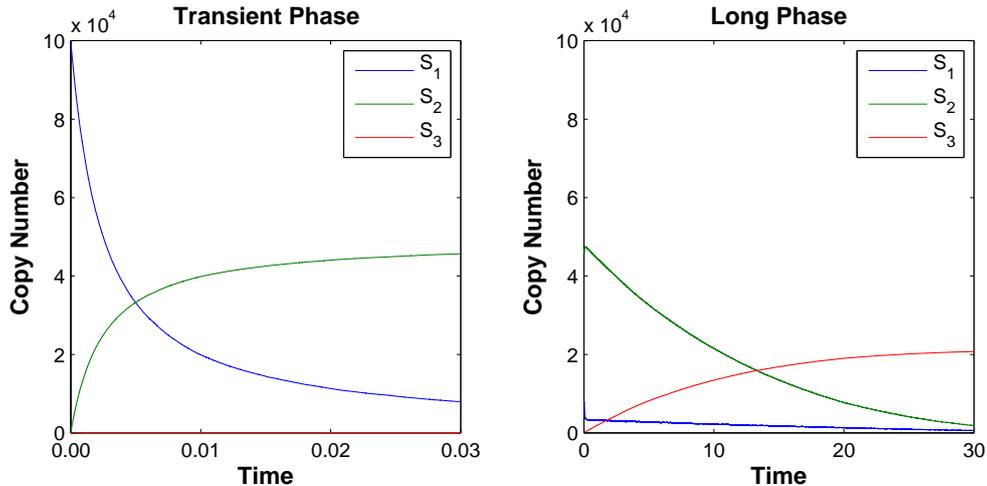

FIG. 1. The temporal evolution of a single sample path of reaction system (12) on two different time-scales. Reaction rates are give in (12) and initial conditions are as described in the text.

We take the initial conditions to be $[X_1, X_2, X_3]^T = [10^5, 0, 0]^T$. Using the Gillespie DM[8], we calculate that the expected population of $S_3$ at time $T = 30$ is $\mathbb{E}[X_3(30)] = 20,591.6 \pm 1.0$. The '$\pm$' term provides a 95% confidence interval for the estimator. This calculation required 36,000 sample paths (this number was chosen on the basis of an initial number of paths which estimated the sample variance), and took a total of 2,089.3 seconds. The desktop computer used to generate results throughout this paper is equipped with a 4.2 GHz AMD FX(tm)-4350 Processor and eight gigabytes of RAM, and code was written in `C++`.

In order to better understand the dynamics of system (12) we consider a typical sample path. In Figure 1 the temporal evolution of a single sample path of the system, generated with the Gillespie DM, is shown on two distinct time scales. A more detailed examination of the trajectory shows that the initial phase is marked by a rush of reaction activity, but once this phase has passed, reaction activity slows dramatically. For this particular realization (figures are broadly similar across all repeats), 613,002 individual reactions were simulated over the time interval $[0, 30]$. Of these, 47,610 were in the first 0.03 seconds. This is equivalent to 1,587,000 reactions per unit time. For the remaining 29.97 units of time, reactions happened at a rate of 19,000 reactions per unit time, which is about 84 times slower.

We now display results of our attempts at applying the fixed time-step multi-level method to this problem. Adopting a trial-and-error approach, we present the simulation times of a range of possible fixed time-step multi-level configurations in Figure 2. In each case, we produce an estimate of $\mathbb{E}[X_3(30)]$ with a 95% confidence interval of semi-length of 1.0. In particular, this demonstrates that the multi-level method can produce an estimate within 652.0 seconds. Each sample in Figure 2 has produced an estimate consistent with the DM method. This therefore represents a factor 3.2 time saving over the Gillespie DM. Whilst significant, this time saving is substantially lower than the results which have been demonstrated elsewhere in the literature for other reaction networks[10,24].

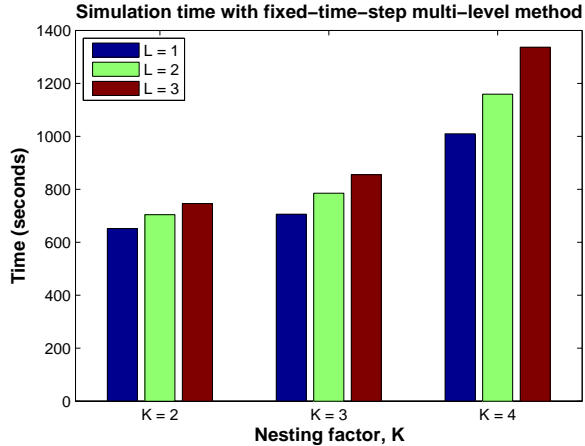

FIG. 2. The simulation time for a range of system configurations for the fixed time-step multi-level method used to estimate $\mathbb{E}[X_3(30)]$ for system (12). Each case uses a different choice of $K$ and $L$; the estimator is unbiased, and we therefore have $L+2$ components to each estimator.

### B. Case Study II: A growth model

We next consider a test model that involves three species. The reaction network is given by:

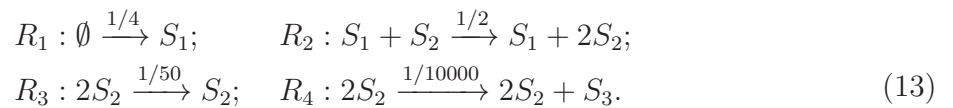

$$R_1 : \emptyset \xrightarrow{1/4} S_1; \qquad R_2 : S_1 + S_2 \xrightarrow{1/2} S_1 + 2S_2;$$
$$R_3 : 2S_2 \xrightarrow{1/50} S_2; \quad R_4 : 2S_2 \xrightarrow{1/10000} 2S_2 + S_3. \tag{13}$$

The reaction rates we will use are indicated in (13) and the initial conditions are $[X_1, X_2, X_3]^T = [1, 5, 0]^T$. In Figure 3 we present a sample trajectory of this system for $t \in [0, 100]$. This time reaction activity increases dramatically over the course of the simulation. Suppose we wish to estimate $\mathbb{E}[X_3(100)]$. The DM provides an estimate of this quantity as $1,535.9 \pm 1.0$. This calculation takes 93,408 seconds (approximately 26 hours) on our PC and uses 1,600,000 sample paths (as with the previous example, a number of preliminary simulations were used to estimate the sample variance, and hence the total number of simulations required). In Figure 4 we show that the fixed time-step multi-level method provides at least a factor 53.9 time saving over the Gillespie DM, as the multi-level method estimates this quantity, with a 95% confidence interval of semi-length 1.0, within 1,733.4 seconds. However, we will demonstrate through the use of an adaptive multi-level algorithm, even this significant saving can be improved upon.

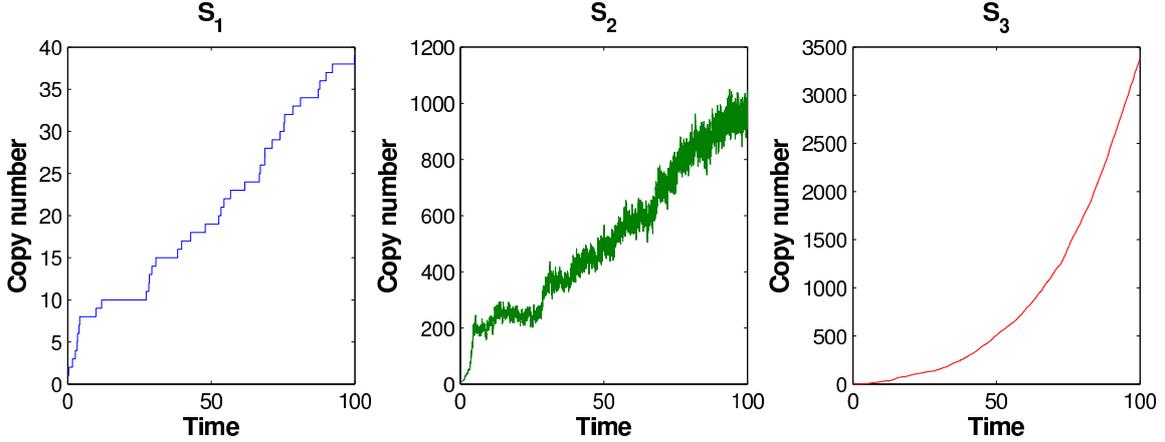

FIG. 3. The temporal evolution of species $S_1$, $S_2$ and $S_3$ which comprise a single sample path of reaction system (13). Reaction rates are give in (13) and initial conditions are as described in the text.

## C. Disadvantages of fixed time-step multi-level

Generating trajectories using the tau-leaping method with a fixed choice of $\tau$ throughout a simulation poses inherent difficulties. Firstly, for temporal regions in which species numbers are changing rapidly we need to be careful not to choose $\tau$ too large that the reaction propensities change considerably over the course of a leap $\tau^9$. At its worst, too large a $\tau$ can render the tau-leap method numerically unstable and therefore non-convergent. With a fixed choice of $\tau$ this means that the temporal region of the path that requires the most stringent bound on $\tau$ determines the overall bound, $\tau_{critical}$. This limits the choices for $\tau_0$ and consequently $\tau_1, \tau_2, \ldots$ in the multi-level method. In particular, it is possible that the time taken to generate a single tau-leaping path with time step $\tau < \tau_{critical}$ is greater than that required generate a single sample path using the DM, rendering the multi-level method redundant.

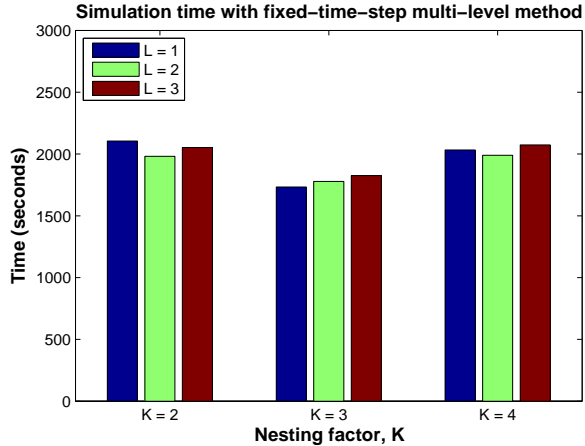

FIG. 4. The simulation time for a range of system configurations for the fixed time-step multi-level method used to estimate $\mathbb{E}[X_3(100)]$ for system (13). Each case uses a different choice of $K$ and $L$; the estimator is unbiased, and we therefore have $L+2$ components to each estimator.

Secondly, at different times during the evolution of a sample path, the reaction propensities will change at different rates. In reaction system (12), within the initial transient phase of a simulation, reaction propensities change quickly with respect to time and therefore must be updated frequently in order to maintain accuracy of the tau-leap method. However, in the slower phase, propensity functions change more slowly and hence larger time-steps can be tolerated between recalculation. Similarly, in reaction system (13), the reaction propensity of $R_4$, the reaction propensity associated with the production of $S_3$, is more sensitive to population changes at higher populations of $S_2$ than at lower populations (as it is proportional to $X_2(X_2 - 1)$).

This means that if we would like to use a fixed number of $\tau$ leaps to generate sample paths over the interval $[0, T]$ to generate a particular estimator, for example, varying the lengths of the leaps over the course of each individual sample path may give rise to a more accurate estimator. As such, we next present a generalization of the multi-level method an adaptive choice of $\tau$. Although our revised method introduces some additional computational overheads it can give rise to significantly reduced simulation times.

## IV. A NEW MULTI-LEVEL APPROACH

The key difference between the original multi-level method[10] described in Section II and our improved approach is that the sample paths generated on each level will no longer be generated using the tau-leap method with constant values of $\tau$. Instead of indexing each set of sample paths by a choice of $\tau_\ell$ we will work with a control parameter, $\xi_\ell$, and the time steps for each set of sample paths will be determined according to a formula parameterized by $\xi_\ell$. As such, we will use a *control parameter ensemble*, $\boldsymbol{\xi} := (\xi_0, \xi_1, \xi_2, \ldots, \xi_L)$, to generate our multi-level estimator. A biased multi-level estimator, $\mathbb{Q}_b$, is given by the telescoping sum of $L+1$ components

$$\mathbb{Q}_b = \mathbb{E}[Z_{h(\xi_L)}] = \mathbb{E}[Z_{h(\xi_0)}] + \mathbb{E}[Z_{h(\xi_1)} - Z_{h(\xi_0)}] + \ldots + \mathbb{E}[Z_{h(\xi_L)} - Z_{h(\xi_{L-1})}], \qquad (14)$$

where $h(\cdot)$ is the function which maps the control parameter to a regime of $\tau$ selection. In addition, an unbiased estimator of $L+2$ terms is given by:

$$\mathbb{Q}_u = \mathbb{Q}_b + \mathbb{E}[X - Z_{h(\xi_L)}], \qquad (15)$$

where $X$ is the value of the random variable of interest generated using an exact SSA.

### A. The base level, $Q_0$

The base level estimator

$$Q_0 := \mathbb{E}[Z_{h(\xi_0)}], \qquad (16)$$

can be estimated with the usual tau-leaping algorithm described in Section II. Details of the algorithm we use to choose $\tau$ in this work are given in Section V A.

## B. The tau-leaping correction terms, $Q_\ell$

We now describe an approach for estimating terms of the form $Q_\ell := \mathbb{E}[Z_{h(\xi_\ell)} - Z_{h(\xi_{\ell-1})}]$, where $\ell \in \{1, 2, \ldots L\}$. We have

$$\begin{aligned} Q_\ell &= \mathbb{E}\left[ Z_{h(\xi_\ell)} - Z_{h(\xi_{\ell-1})} \right] \\ &\approx \frac{1}{n_\ell} \sum_{r=0}^{n_\ell} \left\{ Z^{(r)}_{h(\xi_\ell)} - Z^{(r)}_{h(\xi_{\ell-1})} \right\}, \end{aligned} \qquad (17)$$

where $Z^{(r)}_{h(\xi_\ell)}$ represents the population of the $i$-th species at a time $T$ in the $r$-th sample path, where tau-leaping with time steps determined according to rule $h(\xi_\ell)$ have been used. If we are able to generate sample paths to estimate (17) so that $Q_\ell$ has a low sample variance, then few sample paths will be required to attain a desired statistical error.

As we are constructing a Monte Carlo estimator, we require each of the bracketed $\left\{ Z^{(r)}_{h(\xi_\ell)} - Z^{(r)}_{h(\xi_{\ell-1})} \right\}$ terms in (17) to be independent of the other bracketed terms in (17). To generate the $r$-th sample value, $\left\{ Z^{(r)}_{h(\xi_\ell)} - Z^{(r)}_{h(\xi_{\ell-1})} \right\}$, we will need to simultaneously generate two sample paths using tau-leaping, but with the reaction propensities updated at different times. In describing our algorithm we do not need to specify a particular method for choosing $\tau$, however, the method used to present results in this work is outlined in Section V A.

As with the fixed time-step multi-level method, the key point to note is that there is no need for $Z^{(r)}_{h(\xi_\ell)}$ and $Z^{(r)}_{h(\xi_{\ell-1})}$, to be independent of one another. This is because our estimator $Q_\ell$ does not depend on the actual copy numbers within each system, but on the difference between these random variates. Hence, $Q_\ell$ can be determined with a Monte Carlo simulator taking samples of the difference only. By recalling that

$$\begin{aligned} \mathrm{Var}[Z_{h(\xi_\ell)} - Z_{h(\xi_{\ell-1})}] &= \mathrm{Var}[Z_{h(\xi_\ell)}] + \mathrm{Var}[Z_{h(\xi_{\ell-1})}] \\ &\quad - 2 \cdot \mathrm{Cov}[Z_{h(\xi_\ell)}, Z_{h(\xi_{\ell-1})}], \end{aligned} \qquad (18)$$

we note it is therefore permissible, and in our interests, for $Z^{(r)}_{h(\xi_\ell)}$ and $Z^{(r)}_{h(\xi_{\ell-1})}$ to exhibit a strong positive correlation.

We achieve this positive correlation in the species of interest, $S_i$, at time $T$ by keeping the state vectors of each system as similar to each other as possible over the time-span of the sample path. Since both systems start with the same initial conditions, one approach is to aim to have each reaction channel fire a similar number of times in both systems so that the corresponding population levels are similar. It is, however, crucial that the distributions of the random variables $Z_{h(\xi_\ell)}^{(r)}$ and $Z_{h(\xi_{\ell-1})}^{(r)}$ are the same as would be produced by the corresponding tau-leaping method with the appropriate control parameter.

We briefly discuss the thickening property of Poisson processes. If Poisson variates $\mathcal{P}_1$, $\mathcal{P}_2$, and $\mathcal{P}_3$ are of parameter $a$, $b$, and $a+b$, respectively, we have $\mathcal{P}_1(a+b) = \mathcal{P}_2(a) + \mathcal{P}_3(b)$. In this case, the equality is in distribution. This means that a Poisson random variate with parameter $a+b$ can be generated by generating two Poisson variates, one with parameter $a$ and the other with parameter $b$, and then summing these together.

We now state our general approach to simultaneously calculating the sample paths $Z_{h(\xi_\ell)}^{(r)}$ and $Z_{h(\xi_{\ell-1})}^{(r)}$, which we will denote the *fine* and *coarse* paths, respectively, up to a terminal time $T$:

1. let $t$ denote the current system time. Initially set it to equal $t_0$;

2. let $\boldsymbol{Z_c}(t)$ and $\boldsymbol{Z_f}(t)$ (resp.) denote the state vectors of the approximate *coarse* and *fine* (resp.) paths, at time $t$, with time steps determined according to $h(\xi_{\ell-1})$ and $h(\xi_\ell)$ (resp.). Set these both to equal the desired initial conditions;

3. for each reaction channel, $R_j$, define $a_j^c$ to be its propensity function when considering the *coarse* resolution path, and, similarly, define $a_j^f$ to be the propensity function for the *fine* path. Based on $\boldsymbol{Z_c}(t)$ and $\boldsymbol{Z_f}(t)$, calculate reaction propensities $a_j^c$ and $a_j^f$ for each reaction channel, $R_j$;

4. calculate the next update times (NUTs), $T_c$ and $T_f$, for the *coarse* and *fine* paths, respectively. Set $T_c := t + h(\xi_{\ell-1}, \boldsymbol{Z_f}(t))$ and $T_f := t + h(\xi_\ell, \boldsymbol{Z_c}(t))$: these represent the times at which the *coarse* and *fine* reaction propensities need to be updated, respectively;

5. set $\eta := \min\{T_c, T_f, T\} - t$. Our aim is to provide a new state vector for both the *coarse* and *fine* paths at time $\min\{T_c, T_f, T\}$. Including $T$ in this minimum ensures that we get an estimated population up until time $T$ but no further. For each reaction channel,

$R_j$, define:

$$\begin{aligned} b_j^1 &= \min\{a_j^f, a_j^c\}, \\ b_j^2 &= a_j^c - b_j^1, \\ b_j^3 &= a_j^f - b_j^1. \end{aligned} \quad (19)$$

We note that by the thickening property, with a time step of $\eta$,

$$\begin{aligned} \mathcal{P}(a_j^c \cdot \eta) &\sim \mathcal{P}(b_j^1 \cdot \eta) + \mathcal{P}(b_j^2 \cdot \eta), \\ \mathcal{P}(a_j^f \cdot \eta) &\sim \mathcal{P}(b_j^1 \cdot \eta) + \mathcal{P}(b_j^3 \cdot \eta); \end{aligned} \quad (20)$$

6. for $r = 1, 2,$ and 3, and reaction channels indexed by $j = 1, \ldots, M$, generate Poisson random numbers, $Y_j^r$, with parameters $b_j^r \cdot \eta$, and provide updated state vectors:

$$\begin{aligned} \boldsymbol{Z_c}(t+\eta) &:= \boldsymbol{Z_c}(t) + \sum_{j=1}^{M}(Y_j^1 + Y_j^2)\boldsymbol{\nu_k}, \\ \boldsymbol{Z_f}(t+\eta) &:= \boldsymbol{Z_f}(t) + \sum_{p=1}^{M}(Y_j^1 + Y_j^3)\boldsymbol{\nu_k}; \end{aligned} \quad (21)$$

7. set $t := t + \eta$ (or, equivalently, $t := \min\{T_c, T_f, T\}$);

8. if $t = T_c$, then, if required, update $a_j^c$ for each reaction channel, $R_j$, using $\boldsymbol{Z_c}(t)$. Similarly, if $t = T_f$ update $a_j^f$ if needs be. In each case where an update occurs, update $T_c$ or $T_f$ to represent the new NUT by setting $T_c := T_c + h(\xi_{\ell-1}, \boldsymbol{Z_c}(t))$ or $T_f := T_f + h(\xi_\ell, \boldsymbol{Z_f}(t_0))$ as required;

9. if $t < T$ return to step five.

Figure 5 provides a visual representation of how the first four iterations of the algorithm might behave. Our algorithm shows that it is possible to decide on the time steps of the *coarse* and *fine* paths independently, but generate them simultaneously. Perhaps most importantly, it is clearly true that this algorithm has no need for the increments of the *fine* path to be nested within (or, indeed, even to be smaller than) those of the *coarse* path.

## C. The final level, $Q_{L+1}^*$

The coupling on the final level to produce the corrector $Q_{L+1}^* := \mathbb{E}[X - Z_{h(\xi_L)}]$ can be done using, for example, a method akin to the Modified Next Reaction Method[25] or Gillespie's

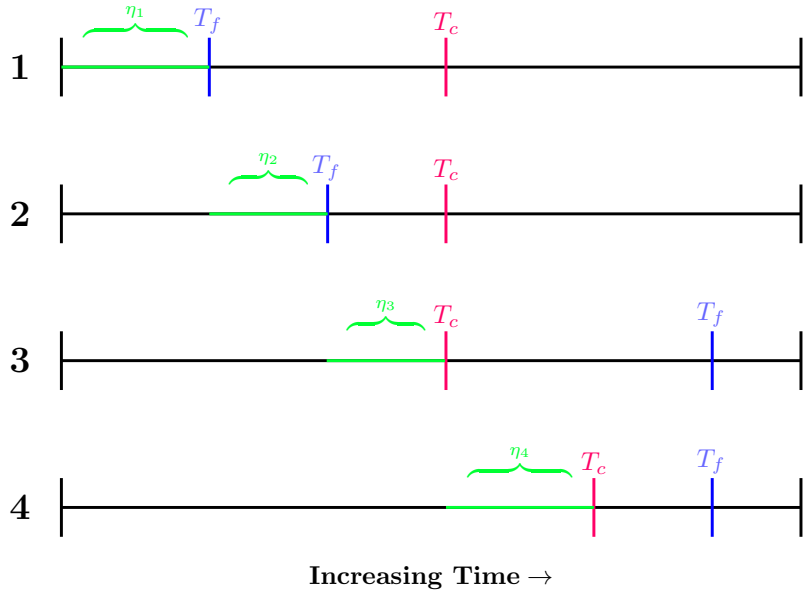

FIG. 5. A diagrammatic representation of a possible first four iterations of the algorithm, shown on a time axis. The vertical lines represent the discretization of time: the NUT of the *fine* system is shown in blue, and the corresponding update times of the *coarse* path are in red. The green shows time steps, $\eta_1, \ldots, \eta_4$, that are used for each iteration. For the first iteration, the NUT of the *fine* path, $T_f$, is reached before the NUT of the *coarse* path, $T_c$. Consequently, the paths are advanced to the *fine* NUT with a time step of $\eta_1$. We update the propensity values of the *fine* path only, and revise the NUT of the *fine* path. The second iteration starts by noting that the NUT of the *fine* path again occurs sooner than the NUT of the *coarse* path, and so a jump of $\eta_2$ to reach the *fine* NUT is implemented. The propensity values and NUT of the *fine* path are updated. For iteration three, $T_f$ is larger than $T_c$, and so a jump of $\eta_3$ is taken. In this case, a new set of propensity values and a new NUT is calculated for the *coarse* path. The fourth iteration progresses the system to $T_c$ (as $T_c < T_f$) and the appropriate updates are performed.

DM[22].

## V. NUMERICAL EXAMPLES

We now return to the motivating examples of Section III and implement an adaptive multi-level method to demonstrate the efficiency of our approach. In each case, results generated using the Gillespie DM and a fixed time-step multi-level implementation are compared. First, however, we outline our method for adaptive choice of $\tau$.

## A. Adaptive choice of $\tau$

The method of adaptive $\tau$ choice we employ in this work is that of Cao, Gillespie and Petzold[19] (the CGP method). It is predicated on the 'leap condition': that is, over a time interval $\tau$, the propensity functions within a system should remain 'approximately constant'. In other words, the change in each propensity function should be small in relation to its magnitude. If the state vector at a time $t$ is given by $\boldsymbol{X}(t) = \boldsymbol{x}$, then denote the change in the propensity function for channel $R_j$ from time $t$ to $t + \tau$ as $\Delta_\tau a_j(\boldsymbol{x})$. In order to achieve the leap condition with high probability we follow the approach of CGP and insist that

$$\tau \leq \frac{\max\{\xi_\ell/g_i \cdot x_i, c_i\}}{|\sum_j \nu_{ij} a_j(x)|} \quad \text{and} \quad \tau \leq \frac{\max\{\xi_\ell/g_i \cdot x_i, c_i\}^2}{\sum_j \nu_{ij}^2 a_j(x)}, \tag{22}$$

for all *reactant* species $S_i$ (i.e. where $X_i$ is an argument of some propensity function). Here $\xi_\ell$ is the control parameter described previously, $g_i$ is a weight function which indicates the relative effect of changes in $X_i$ on the leap condition and $c_i$ is the minimum expected change allowed: it is normally set to unity as this is the minimum possible change in population. In the original CGP method, special care is taken to mitigate the risk of a negative population being realized, and certain reaction channels are labeled as 'critical' and afforded special treatment. We temporarily overlook this complication and instead take

$$\tau = \min_{i \in I_r} \left\{ \frac{\max\{\xi_\ell/g_i \cdot x_i, c_i\}}{|\sum_j \nu_{ij} a_j(x)|}, \frac{\max\{\xi_\ell/g_i \cdot x_i, c_i\}^2}{\sum_j \nu_{ij}^2 a_j(x)} \right\}, \tag{23}$$

with the parameters as previously defined, and $I_r$ as the set of reactant species.

### 1. Case Study I: A dimerization model

We first consider the reaction network (12). In Table I the results of estimating $\mathbb{E}[X_3(30)]$ using a range of adaptive multi-level configurations are given. These show unbiased estimates, according to equation (15). In each case, an ensemble of control parameters is given; these parameters are used to implement our chosen method[19] of adaptively selecting $\tau$. We used the method outlined in Anderson and Higham[10] to determine how many sample paths to generate on each level.

| Control parameter ensemble | Estimate | Time |
|---|---|---|
| (0.16, 0.08, 0.04, 0.02) | $20,590.8 \pm 1.0$ | 86.0 s |
| (0.18, 0.06, 0.02) | $20,592.4 \pm 1.0$ | 72.6 s |
| (0.32, 0.08, 0.02) | $20,591.5 \pm 1.0$ | 100.8 s |
| (0.10, 0.02) | $20,591.5 \pm 1.0$ | 82.8 s |
| (0.12, 0.02) | $20,591.2 \pm 1.0$ | 76.3 s |

TABLE I. A range of control parameter ensembles are tested in order to estimate $\mathbb{E}[X_3(30)]$ for reaction system (12).

| Level | $(\xi_\ell, \xi_{\ell-1})$ | Mean | Variance | Samples | Time |
|---|---|---|---|---|---|
| 0 | 0.18 | 20,699.8 | 9,566.7 | 77,109 | 32.5 s |
| 1 | (0.06, 0.18) | -88.9 | 169.5 | 4,543 | 11.0 s |
| 2 | (0.02, 0.0.06) | -16.0 | 45.1 | 1,163 | 11.5 s |
| 3 | (DM, 0.02) | -2.6 | 15.0 | 242 | 17.6 s |
| **Total:** | | **$20,592.4 \pm 1.0$** | | - | **72.6 s** |

TABLE II. The contribution of each term in equation (15) to the overall multi-level estimator of $\mathbb{E}[X_3(30)]$ for reaction system (12) with $\boldsymbol{\xi} = (0.18, 0.06, 0.02)$.

Most impressively, for control parameters $\boldsymbol{\xi} = (0.18, 0.06, 0.02)$ we achieve an estimate of $\mathbb{E}[X_3(30)]$ as $20,592.4 \pm 1.0$ within 72.6 seconds. This is approximately 28.8 times faster than the DM method. In addition, it is 9.0 times faster than the most efficient results presented in Figure 2 where the fixed time-step multi-level method was employed. In Table II we show the contribution from each of the four levels to this estimator.

### 2. Case Study II: A growth model

The motivating example of interest is reaction network (13). In order to demonstrate the efficacy of our adaptive multi-level method in this particular case, a number of algorithm configurations have been tested and the results given in Table III. For the control parameters $\boldsymbol{\xi} = (1, 0.2, 0.04)$, calculation of $\mathbb{E}[X_3(100)]$ was performed in 621.9 seconds, giving an estimator value of $1,535.3 \pm 1.0$, which is comparable with that given by the Gillespie DM in Section III (which took a factor of 150.2 times longer to generate). In addition, our calculation was completed 2.8 times quicker that the most efficient configuration of the fixed time-step multi-level approach we found. In Table IV we show the contribution from each of the four levels to this estimator.

| Control parameter ensemble | Estimate | Time |
|---|---|---|
| $(1, 0.5, 0.25, 0.125, 0.0625)$ | $1,535.3 \pm 1.0$ | 687.7 s |
| $(0.9, 0.3, 0.1)$ | $1,535.2 \pm 1.0$ | 630.4 s |
| $(0.9, 0.3, 0.1, 0.03)$ | $1,536.5 \pm 1.0$ | 648.4 s |
| $(0.8, 0.2, 0.05)$ | $1,536.6 \pm 1.0$ | 638.0 s |
| $(1, 0.2, 0.04)$ | $1,535.3 \pm 1.0$ | 621.9 s |

TABLE III. A range of control parameter ensembles are tested out as a multi-level configurations in order to estimate $\mathbb{E}[X_3(100)]$ for reaction system (13).

| Level | $(\xi_\ell, \xi_{\ell-1})$ | Mean | Variance | Samples | Time |
|---|---|---|---|---|---|
| 0 | 1.0 | 1,433.6 | 355,662.0 | $2.09 \times 10^6$ | 395.5 s |
| 1 | (0.2, 1.0) | 93.3 | 1,743.9 | 57,941 | 80.9 s |
| 2 | (0.04, 0.2) | 7.8 | 162.8 | 5,868 | 65.5 s |
| 3 | (DM, 0.04) | 0.6 | 39.1 | 1,129 | 80.0 s |
| **Total:** | | **$1,535.3 \pm 1.0$** | | - | **621.9 s** |

TABLE IV. The contribution of each term in equation (15) to the overall multi-level estimator of $\mathbb{E}[X_3(100)]$ for reaction system (13) with $\boldsymbol{\xi} = (1.0, 0.2, 0.04)$.

## VI. DISCUSSION

The multi-level method provides impressive time savings by combining a number of SSAs in an efficient manner to generate system statistics of interest. However, the original formula-tion of the algorithm required each sample path to be generated using a fixed value of $\tau$, and each of the levels to be nested, in the sense that $\tau_\ell = \tau_{\ell-1}/K$ where $K \in \{2, 3 \ldots\}$. In this work we have shown how to extend the multi-level method to remove these restrictions, and hence make it applicable to the study of systems where reaction activity varies substantially on the timescale of interest. We have demonstrated the efficiency of our method using two example systems, and in each case used the CGP method to define the adaptive choice of $\tau$. However, our algorithm is general in the sense that it can accommodate almost any method for choosing $\tau$ adaptively. Our future work will be directed towards further exploration of efficient methods for adaptively choosing $\tau$ along each sample path and the construction of adaptive hybrid multi-level methods which allow one to switch between using approximate and exact SSAs within the course of a single sample path[24].